\documentclass[11pt]{article}
\usepackage{blindtext}
\pdfoutput=1
\usepackage{subcaption}
\usepackage{graphicx}
\title{A Flap-Type Seabed-Mounted Wave Energy Converter With Hydraulic PTO: Analytical Investigation \& Experimental Analysis}
\date{}
\author{Masoud Masoumi \\ \\ \footnotesize Mechanical Engineering Department, Manhattan College, Riverdale, NY 10471 \\ \footnotesize mmasoumi01@manhattan.edu}

\begin{document}

\maketitle

\textbf{Abstract}\\
This paper provides a study of a flap-type wave energy converter with a hydraulic power-take-off (PTO) system. This study includes the derivation of its equation of motion along with an analysis of natural frequencies of the device. A prototype was built using fillable cylinders and a series of experiments were performed in a wave flume. In these experiments, the pressure values of the hydraulic circuit were changed to investigate the optimal value for the system's pressure for a range of wave periods. The average wave-to-fluid efficiency (of about 15\%) was achieved for a broad range of wave periods. It was also found that a 26\% wave-to-fluid efficiency can be reached for waves with a period of 6.71~s. Based on the results of this analytical study, this device can be further optimized by filling its compartments with seawater.
%\end{abstract}

\providecommand{\keywords}[1]{\textbf{\textit{Keywords}} #1}


\section{Introduction}

The development of economically competitive energy converters to harness clean sources of energy (e.g., ocean waves) has always been a desirable goal for researchers. As one of the few untapped resources of renewable energy, ocean waves have the potential to produce 2.7~TW power. In the United States alone, about 44\% of the available wave energy is recoverable~\cite{pecher2017handbook}. It is also estimated that the west coast of the continental U.S. has waves with an energy level as high as 32 kW/m~\cite{beyene2007digital}. These estimations provide an incentive to develop technologies capable of efficiently converting ocean waves into electricity ~\cite{jacobson2011mapping,isaacs1980ocean}.
Various technologies have been introduced and have undergone rigorous experimental tests and simulations over the past few decades~\cite{sabzehgar2009review, prakash2016wave}. Rusu and Onea~\cite{rusu2018review} provided a review of all available wave energy converter technologies as well as updates on their current state of development. Further, reviews on wave energy harnessing technologies and their future usage in the U.S. as well as globally have been provided by Lehmann et al~\cite{lehmann2017ocean} and Melikoglu~\cite{melikoglu2018current}. The possibility of using ocean wave energy to power ocean robots have also been reviewed by Tian and Yu~\cite{tian2019current}.

Flap-type ocean wave energy converters were developed as the latest wave energy harvesting technology, used to improve the efficiency and the competitiveness of ocean wave energy converters~\cite{tornqvist1978theoretical}. Currently, the two important flap-type seabed-mounted ocean wave energy converters are called the Oyster and the  WaveRoller~\cite{whittaker2007development,cameron2010design, lucas2012development}. Width et al~\cite{windt2014development} implemented a hybrid approach using a finite element method and wave tank testing to study the structural loads on the flap. Finally, Chang et al~\cite{chang2015theoretical} employed a mesh-free technique called Smoothed Particle Hydrodynamics (SPH) for the simulations of their flap-type wave converter.

Henry et al~\cite{henry2013characteristics} used a high-speed camera to investigate the interactions between the ocean waves and the flap-type seabed-mounted ocean wave energy harvesters. They used both numerical methods and experimental testing in this case. Building on prior work, Kumawat et al~\cite{kumawat2019wave} utilized a three-dimensional boundary element method to find the hydrodynamic coefficients of multiple flap-type seabed-mounted wave energy converters. In another study, they investigated the possibility of using a  flap-type harvester along with a wind turbine. They considered three elliptical shape wave energy converters and a 5~MW floating turbine on a platform~\cite{kumawat2019numerical}.

This paper addresses a bottom-hinged oscillating plate wave energy converter with fillable compartments and a hydraulic power take-off (PTO) system. The purpose of this paper is to provide a practical dynamic model for this device and to present the results of these experimental tests for power output estimation. The fundamental working principle is similar to the Oyster with a flap structure, which is hinged at the seabed~\cite{whittaker2012nearshore}. The focus of this study relies on the possibility of manipulating the device's natural frequency using the sea water and to conduct an investigation of the efficient pressure values for its hydraulic system. First, the motion of converter is analytically examined and effective parameters on resonant frequency are investigated. Then, a series of experimental tests are conducted to find the best oil pressure values for the hydraulic PTO mechanism with a cylinder-piston configuration.

\section{Analytical Modeling}
\begin{figure}[h!]
	\centering
	\includegraphics[scale=0.85]{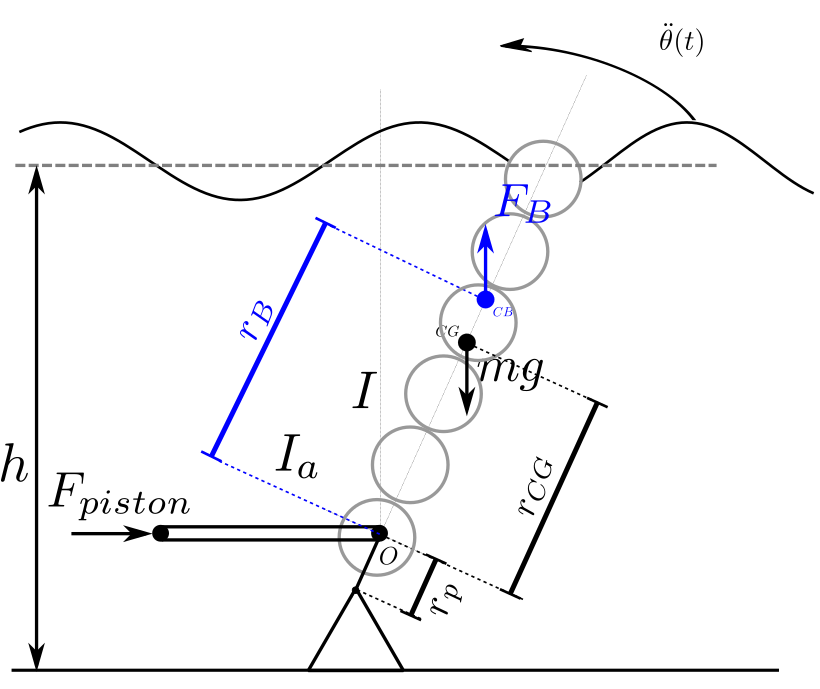}
	\caption{Free body diagram of the device showing hydrodynamic and hydrostatic forces.}
	\label{fig:FBD}
\end{figure}

A schematic representation of the flap-type wave energy converter designed for shallow and intermediate water depth is shown in Figure~\ref{fig:FBD}. The flap is built using six fillable tubes. Two linear hydraulic cylinders are used to convert wave energy from the surge motion into fluid power (see Figure~\ref{fig:FBD}). The general equation of motion can be found using the rotational equivalent of Newton’s second law of motion around pivotal point O (see Figure~\ref{fig:FBD} for the free body diagram)

\begin{equation} \label{eq:eq1}
\big(I+I_a \big)\ddot{\theta}(t) + C_r\dot{\theta}(t)+K\theta(t) = T_{exc}-T_{pto}
\end{equation}
In this equation, drag, wave drift, current and
moorings are disregarded for simplicity and the applied load is a
combination of wave excitation and loads from the machinery\cite{pecher2017handbook}. Further, $I$ and $I_a$ are the flap and added moment of inertia respectively, $C_r$ is the value for radiation damping, $K$ represents stiffness coefficient, $T_{exc}$ is the excitation torque caused by the surge motion of the waves, and $T_{pto}$ is the counter-torque created by the PTO mechanism. The excitation torque is caused by a summation of incident and scattered waves, while added inertia relates to the force caused by the mass of water. Radiation damping is the representative of power exchanged between the sea and the body. Here, both buoyancy force and gravity are as considered hydrostatic forces. 

\begin{figure}[h!]
	\centering
	\includegraphics[scale=0.5]{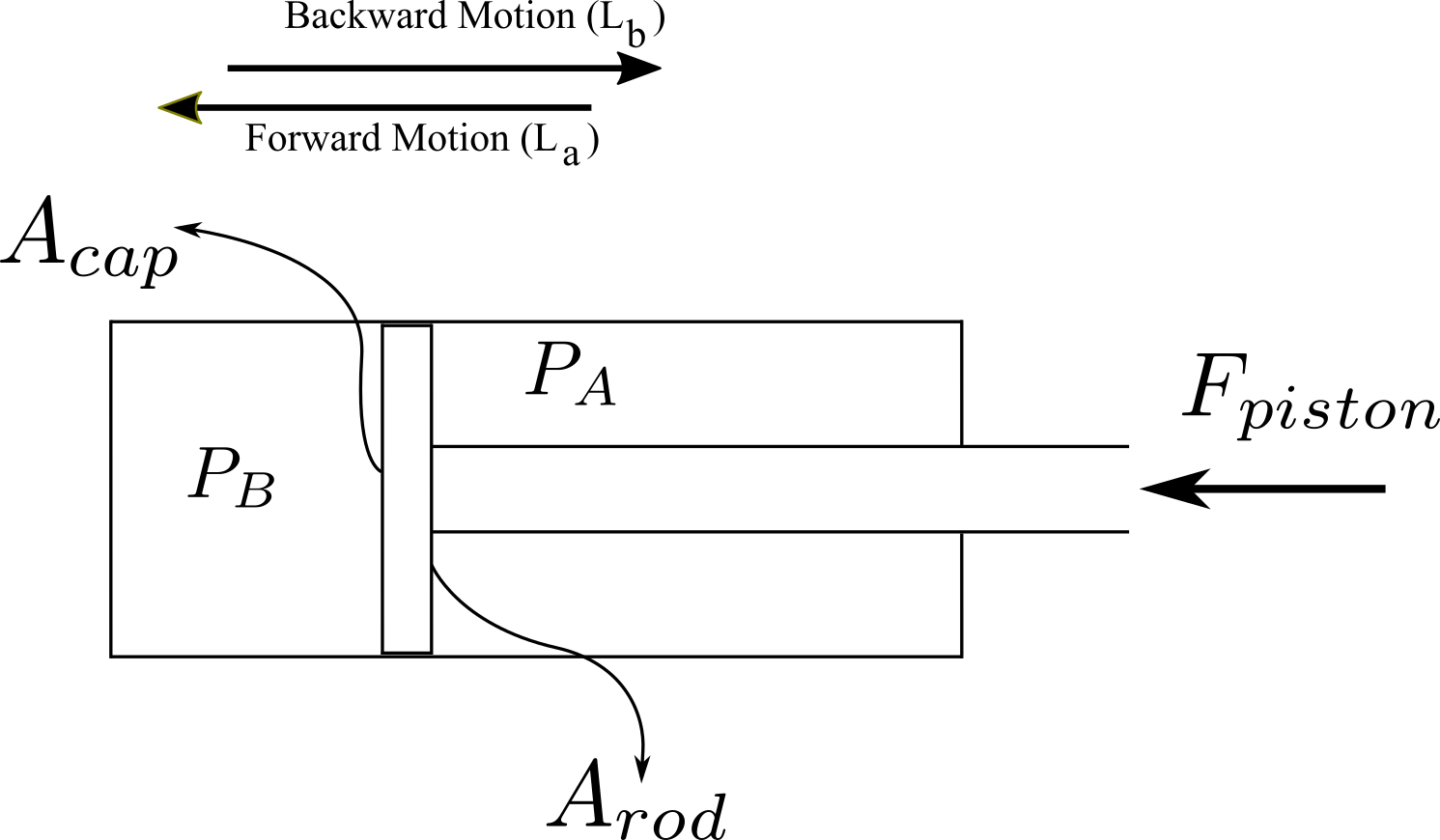}
	\caption{Cylinder-piston configuration as part of the hydraulic PTO and the related parameters.}
	\label{fig:CylinderPiston}
\end{figure}

The PTO mechanism is a hydraulic system with cylinder-piston gadgets on both sides of the oscillating flap. The equation of motion can be further developed by replacing the force with pressure values and areas in the cylinder-piston configuration, as shown in Figure~\ref{fig:CylinderPiston}. The piston force is a function of pressure values, $P_A$ and $P_B$, and areas on both sides of the piston inside the cylinder, $A_{cap}$ and $A_{rod}$. The force and torque from the piston can be found using the pressure values, areas on both sides of the cylinder, and the location of the connecting rod.

\begin{equation} \label{eq:eq2}
F_{piston}=P_BA_{cap}-P_AA_{rod} \qquad , \qquad
T_{pto} = r_p \times F_{piston} 
\end{equation}

\begin{equation} \label{eq:eq4}
T_{pto} = r_p \left( P_BA_{cap}-P_AA_{rod}\right)
\end{equation} 

\begin{figure}[h!]
	\centering
	\includegraphics[scale=0.43]{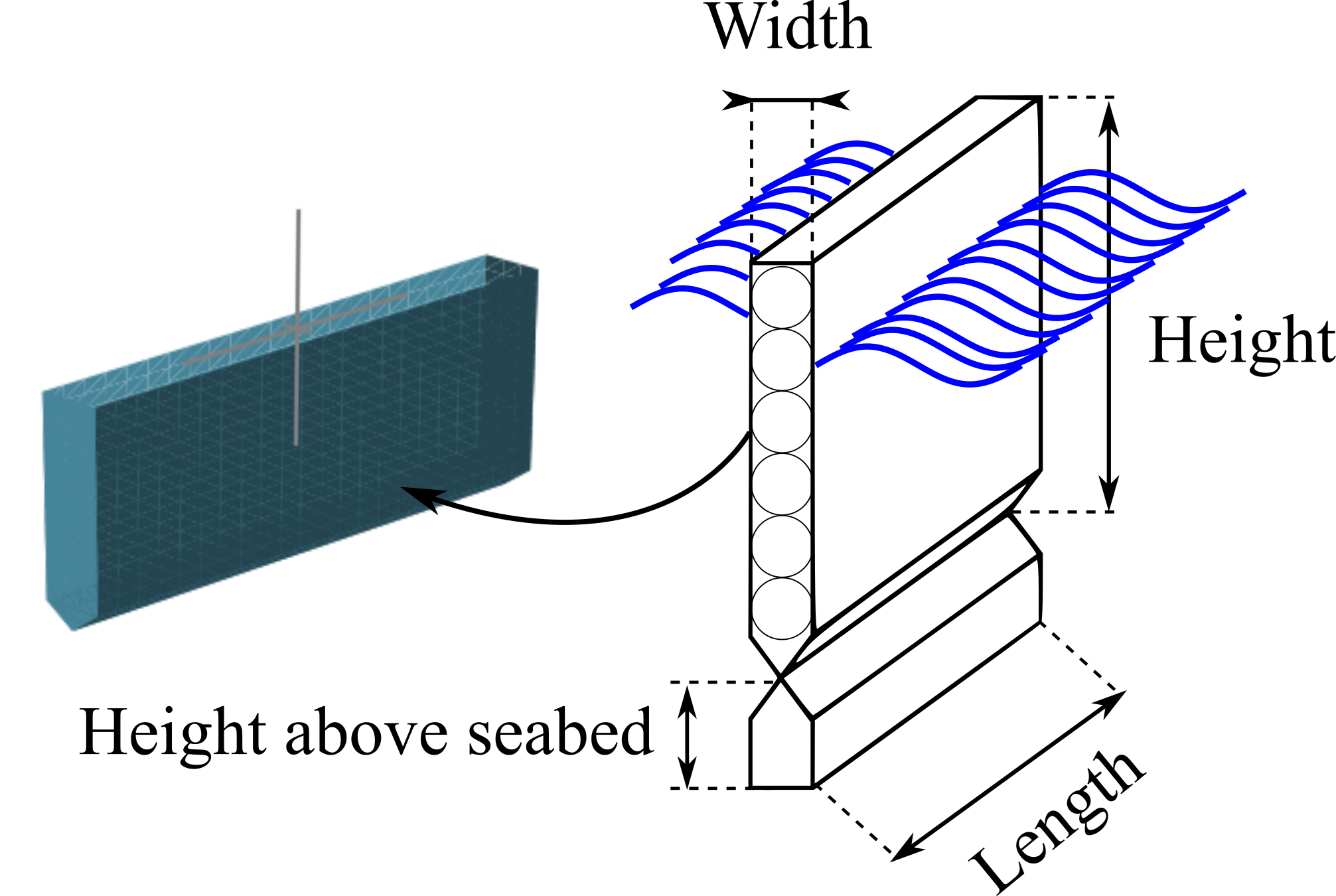}
	\caption{Geometry of the flap as a seabed-mounted wall with a rectangular cross-section along with the corresponding mesh.}
	\label{fig:FlapDimMesh}
\end{figure}

The final expression for the equation of motion can then be represented by substituting these variables into equation~\ref{eq:eq1}, as follows:

\begin{equation} \label{eq:eq5}
\big(I+I_a\big)\ddot{\theta}(t)+C_r\dot{\theta}(t) +K\theta(t) = T_{exc} - r_p \left( P_BA_{cap}-P_AA_{rod}\right)
\end{equation}
\\
Both radiation damping and added inertia are dependent on the frequency of oscillation, as indicated by the frequency argument in the equation of motion given above~\cite{pecher2017handbook}. These values can be calculated using the boundary element method (BEM) for different body shapes interacting with water~\cite{penalba2017using}. The prototype shown in Figure~\ref{fig:FBD} has six hollow PVC cylinders, each with the inner diameter of 10.16~cm (4~in) and outer diameter of 11.43~cm (4.5~in). Two steel bars on one side are used to support the cylinders with a 8.89~cm (3.75~in) distance between the rotating axis and the center of the lowest cylinder. When each cylinder is completely filled with water, it weighs 8 Kg (17.637 lbs). Here, the geometry of the flap is simplified as a seabed mounted flap with a rectangular cross-section, as shown in figure~\ref{fig:FlapDimMesh} with the dimensions listed in Table~\ref{table:table1}.

\begin{table}[h!]
	\centering
	\begin{tabular}{cccc}
		\hline
		Width (m)& Height (m)& Length (m)& Water Depth (m) \\
		\hline
		0.1143 & 0.6858 & 1.03024 & 0.9144 \\
		\hline 
	\end{tabular}
	\caption{Prototype properties} \label{table:table1}
\end{table}

\begin{figure}[h!]
	\centering
	\includegraphics[scale=0.45]{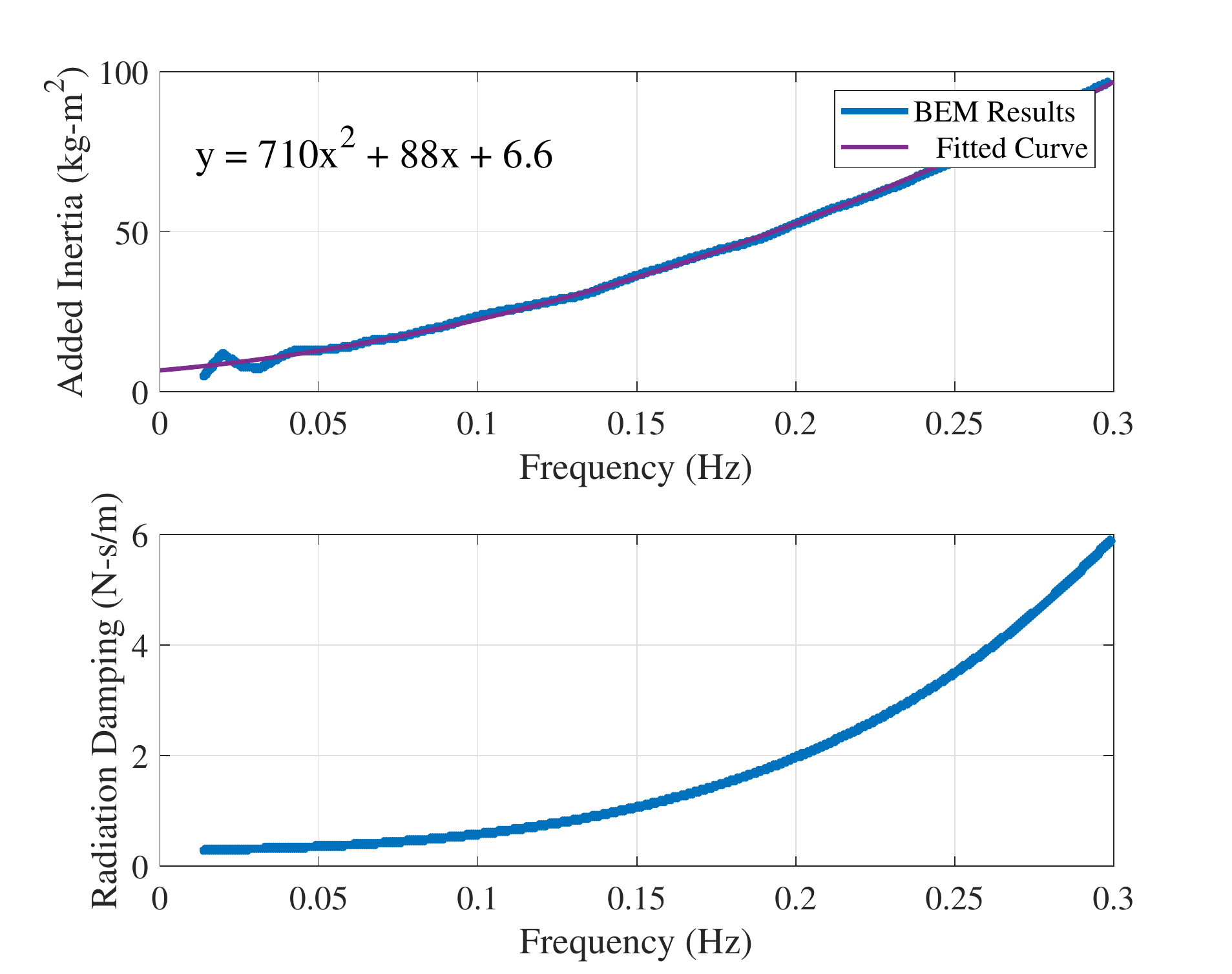}
	\caption{Added inertia and radiation damping for the flap with the geometry defined in table~\ref{table:table1} for the frequency range of 0.02-3~Hz.}
	\label{fig:InertiaDamping}
\end{figure}

BEM was used to calculated the values of added inertial and radiation damping for this flap in a frequency range of 0.02-3~Hz using the approach described in~\cite{verbrugghe2018coupling}. These values are shown in figure~\ref{fig:InertiaDamping}. As can be seen, both added inertia and radiation damping increases when the frequency of oscillation increases. A quadratic equation can be fitted over the values of added inertia in this frequency range with the equation $709.4x^2+87.948x+6.63$, where $x$ represents the frequency in Hz. For this curve fitting, the normal of residuals is 32.772. 

Resonant frequency of the harvester can also be calculated from equation~\ref{eq:eq5} or equation~\ref{eq:eq1} as

\begin{equation} \label{Eqn:NF1}
\omega_n=\sqrt{\frac{K}{I+I_a(\omega)}}
\end{equation}

\begin{figure}[h!]
	\centering
	\begin{subfigure}{0.5\textwidth}
		\centering
		\includegraphics[scale=0.55]{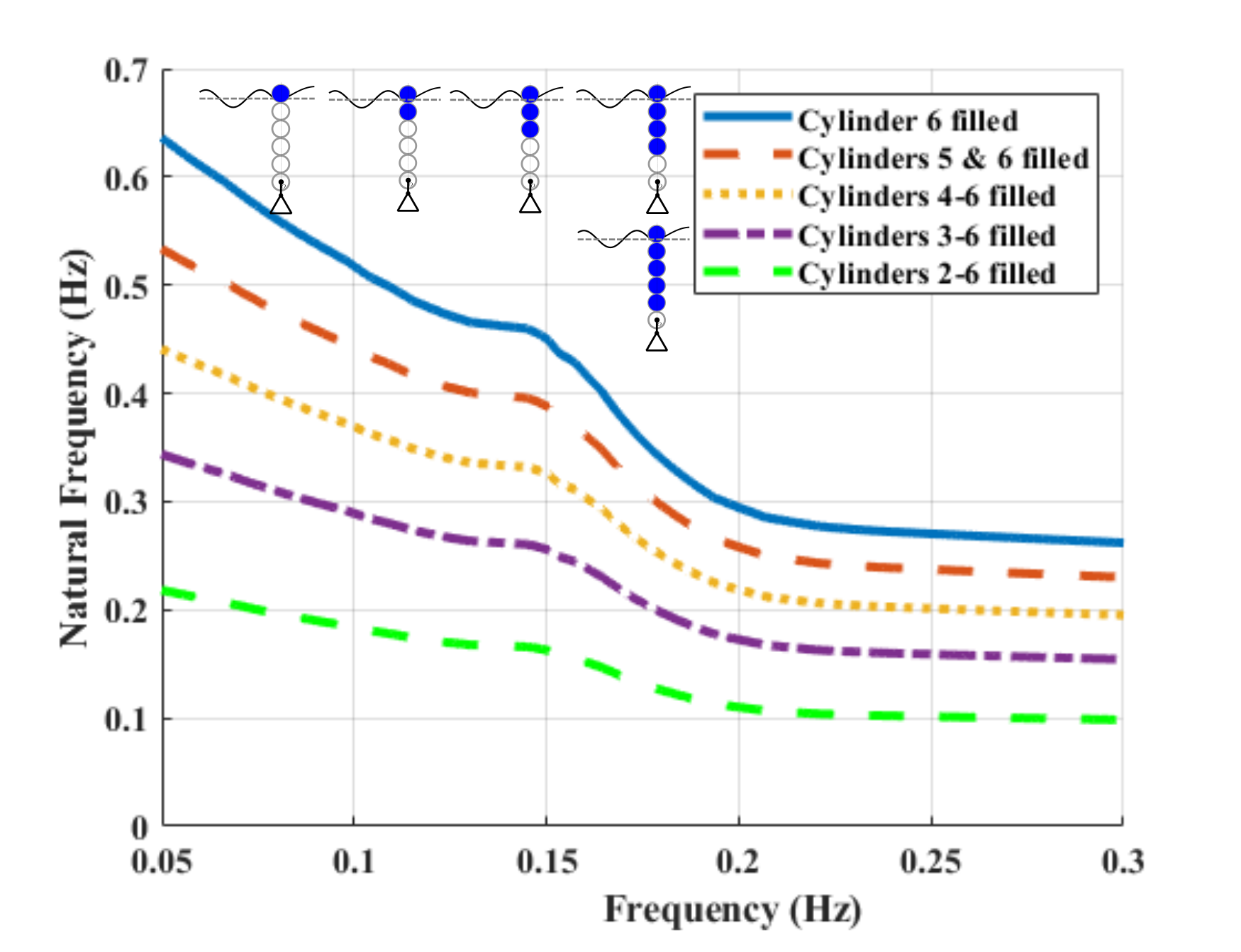}
		\caption{}
		\label{}
	\end{subfigure}
	\begin{subfigure}{0.5\textwidth}
		\centering
		\includegraphics[scale=0.55]{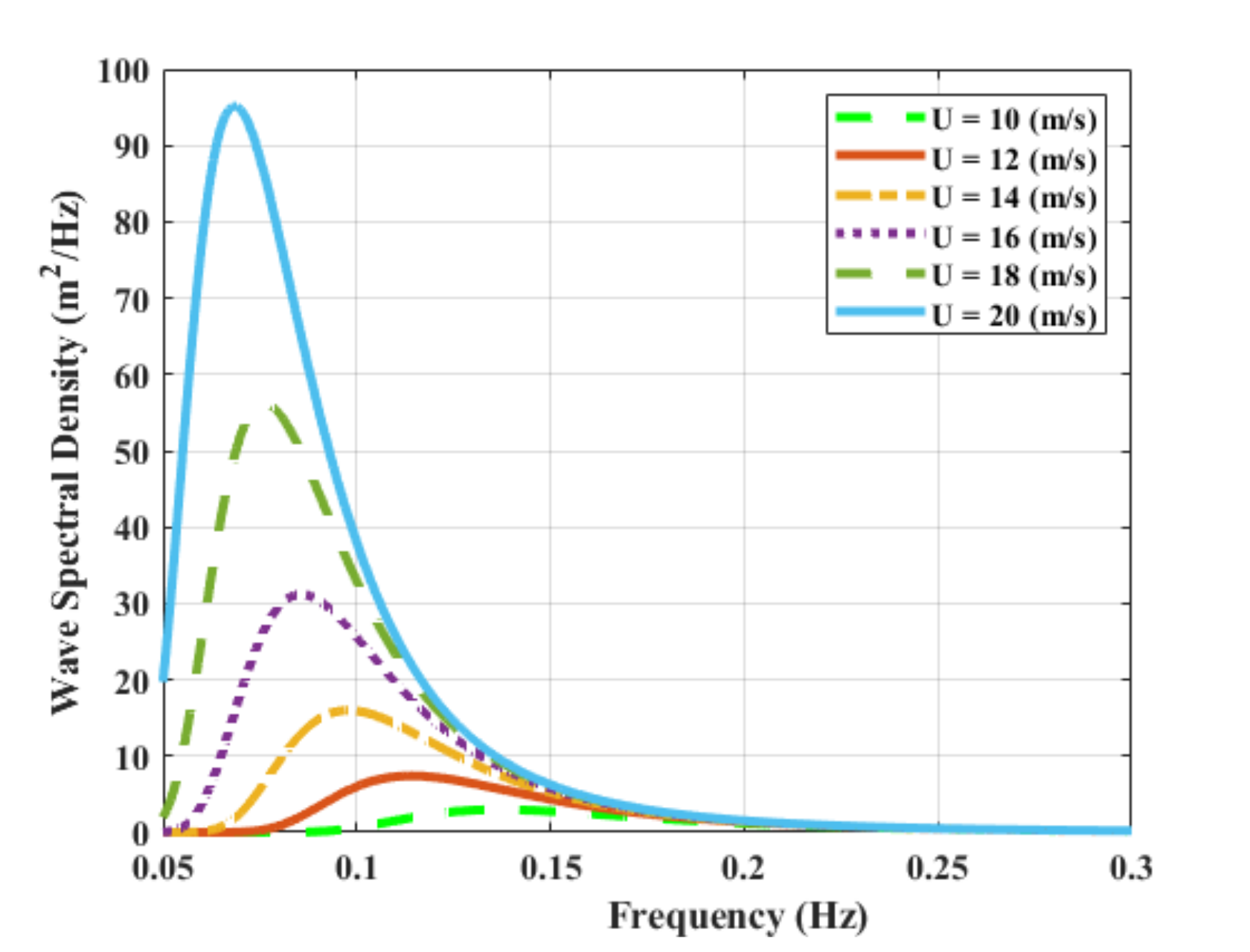}
		\caption{}
		\label{}
	\end{subfigure}
	\caption{(a)~Natural frequencies for five different cases with various mass distributions,and (b)~the wave spectral density for a fully-developed sea-state for different wind speed (U) values.}
	\label{fig:PowerOutputTankBTP15}
\end{figure}

To study the effects of mass and its distribution on natural frequency of the device, five different cases were considered. Case one gauges the impact of the top cylinder filling with water. In the second case, the top two cylinders were filled with water and so on. For each case, added inertia was taken into account using the curve fitted to the values of added inertia shown in figure~\ref{fig:InertiaDamping}. The results along with the schematics of these cases are shown in Figure~\ref{fig:PowerOutputTankBTP15}~(a). Here, the flap was approximated as a wall with a rectangular cross-section. The value $K$ was then estimated using $K=\rho_wgt^3 d/12+(\rho_w -\rho_f)gtdh^2/2$, where $\rho_w$ and $\rho_f$ are the water and device densities, $t$ is the thickness of the flap, $h$ is the water depth, $d$ stands for the width of the flap, and $g$ represents the acceleration of gravity~\cite{zhao2013numerical}. To better gauge the performance of a full-scale device, results were scaled for a $1/20^{th}$ scale machine using Froude’s law since the inertia forces are predominant. Also, the spectra of the fully-developed sea-state was simulated using Pierson-Moscowitz model~\cite{pierson1964proposed} as shown in figure~\ref{fig:PowerOutputTankBTP15}~(b) for different wind speed values, $U$.

As seen in figure~\ref{fig:PowerOutputTankBTP15}, the natural frequency decreases, as the flap weight increases. More specifically, the change in natural frequency for each case asymptotically decreases to a constant value for oscillations with frequencies above 0.2~Hz. There is also a range for a wave's excitation from 0.15~Hz to 0.2~Hz where a sudden change in natural frequency is observed. This is the range in which a relatively small variation in the waves' dominant frequency can greatly affect the system's natural frequencies. This phenomenon is clearer for lighter cases where the mass is primarily concentrated at the top and away from the hinge. Even for the case with five cylinders filled with water, there still exists a relatively large difference between the desired natural frequency, which is below 0.1~Hz, and the harvester's frequency. In other words, the lowest natural frequency that can be achieved using this configuration and in the heaviest case, is far from the frequency of high energy waves. The final configuration of the device depends on the installation site, which determines the dominant wave frequency.

\section{Experimental Study}
The primary aims for this experimental analysis is to study the effects of oil pressure on the harvester's performance and to determine the optimal oil pressure ratio for the hydraulic PTO system. The flap-type bottom-mounted harvester has properties described in Table~\ref{table:table1}. All the experiments were conducted on the flap with no filling and only the oil pressures and waves' time periods were changed during the experiments to study the outcomes.

\subsection{Experiment Setup}
\begin{figure}[h!]
	\centering
	\includegraphics[scale=0.22]{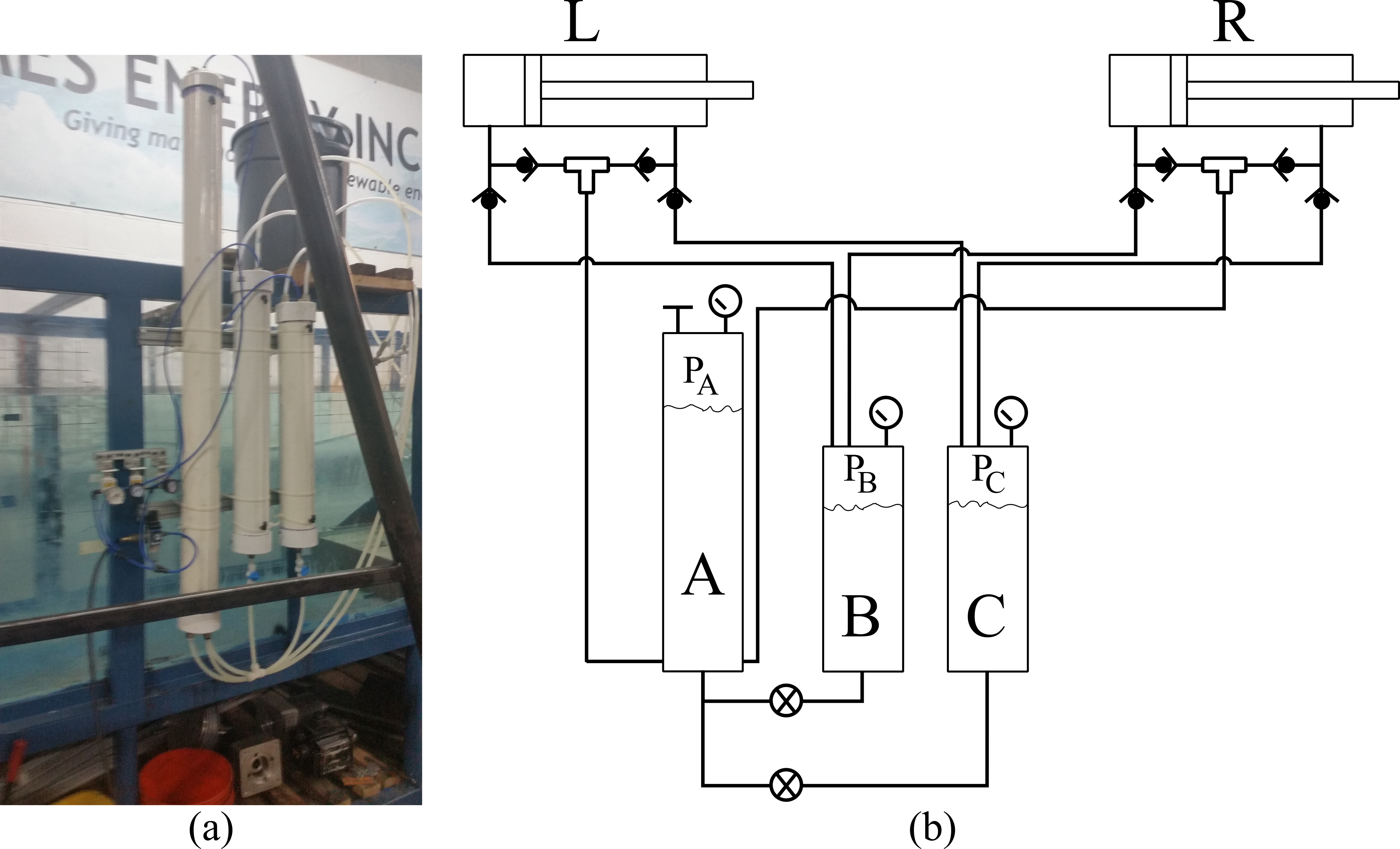}
	\caption{(a) Three tanks installed next to the wave flume, and (b) a schematic diagram of the hydraulic circuit connecting the tanks to cylinder-piston configuration on the flap.}
	\label{fig:HydraulicCircuit}
\end{figure}

The hydraulic circuit setup used for these experiments is shown in figure~\ref{fig:HydraulicCircuit}. To control the oil pressure, three tanks with pressurized oil were used. Tank A is the reservoir and its pressure was always kept constant at 34.474~kPa (5~psi). Tanks B and C control the oil pressure for two different motions in the system. When the flap is moving in the same direction as incident waves, the pistons connected to both sides of the flap move from left to right. The pressure on the left side of each cylinder that primarily controls the damping for this motion is set using tank B. When the flap is moving in the opposite direction of the incident waves, going back towards its initial position, the damping force is principally controlled by the oil pressure in tank C due to the PTO mechanism. The letters $L$ and $R$ are used to indicate the left and right cylinders of the flap in figure~\ref{fig:HydraulicCircuit}~(b). The wave flume used for these experiments is designed and built in house at Brimes Energy. This flume is capable of making waves up to 25.4~cm (10~in) in height and has the size of 121.92~cm (48~in) by 121.92~cm (48~in) with the capacity of approximately 18927~liters (5000~Gallons). The wave generator is controlled by a 7.5~kW servo motor to generate harmonic waves with a 1~s to 3.5~s time period (TP)~\cite{masoumiexperimental} using the in-house LabView program written specifically to generate these waves. Four in-house developed wave gauges with 800~mm probes were used during the experimental testing. Two of these gauges were mounted between the prototype and the wave maker and the other two gauges were between the shore and the prototype.

\begin{figure}[h!]
	\centering
	\begin{subfigure}[b]{0.5\textwidth}
		\centering
		\includegraphics[scale=0.4]{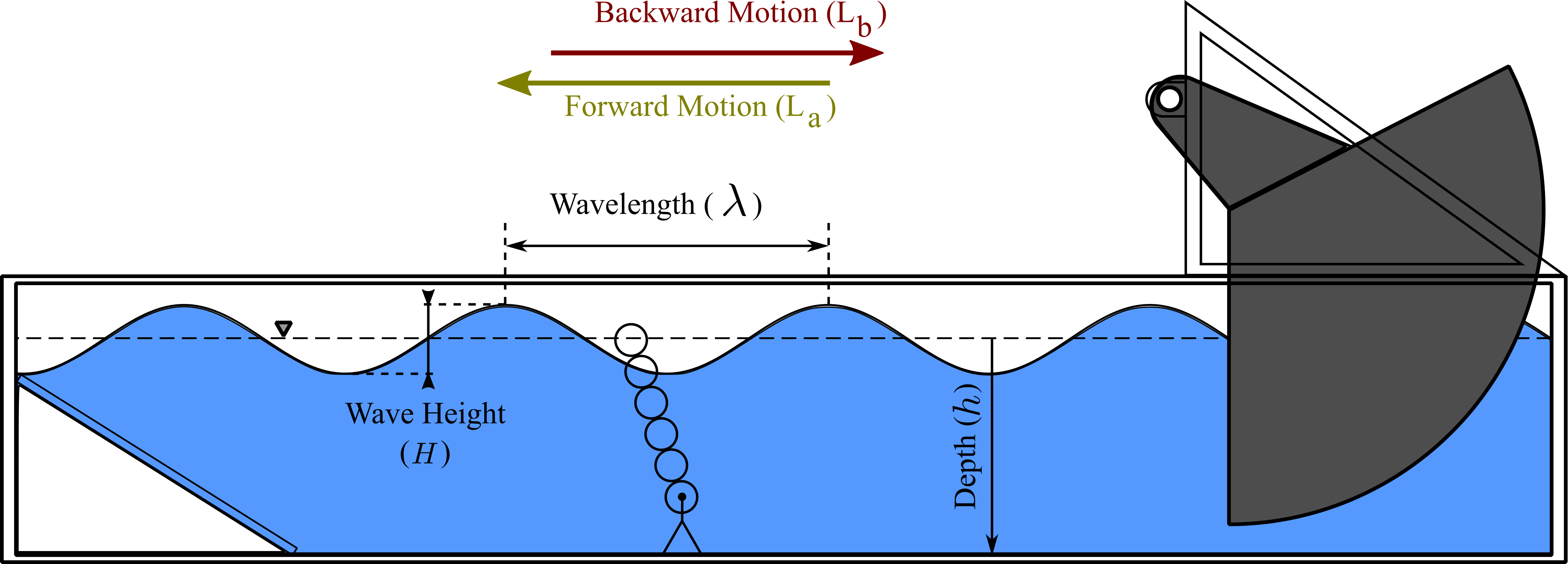}
		\caption{\centering}
	\end{subfigure}
	\begin{subfigure}[b]{0.5\textwidth}
		\centering
		\includegraphics[scale=0.4]{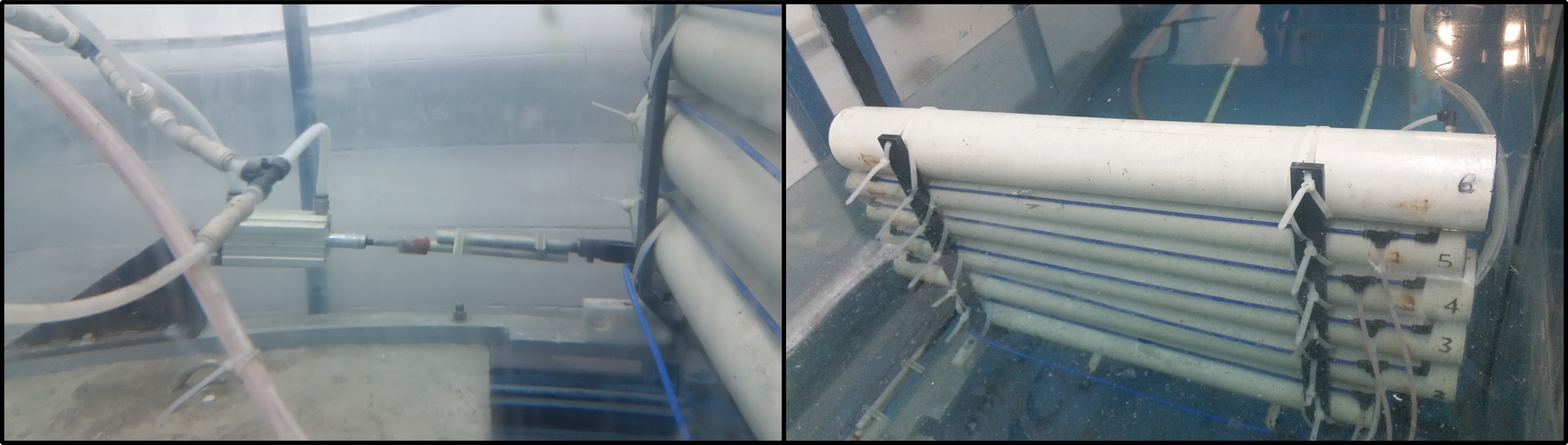}
		\caption{\centering}
	\end{subfigure}
	\caption{(a) A schematic diagram of the wave flume with the prototype inside (not to scale), and (b) the prototype inside the wave flume with connections to hydraulic PTO.}
	\label{fig:WaveTank}
\end{figure}

The flap was hinged at the bottom of the wave flume to a 78.74 $\times$ 101.6 $\times$ 17.78~cm$^3$ (31$\times$40$\times$7~in$^3$) concrete plate. The water depth was kept at 91.44~cm (36~in) throughout all tests and the waves had a height of 15.24~cm (6~in). Wave periods varied from 1~s to 3~s with 0.5~s steps. For each test, 25 wave cycles were used. Pressure in tank B and tank C varied from 68.948~kPa (10~psi) to 413.685~kPa (60~psi) and 137.895~kPa (20~psi) to 413.685~kPa(60~psi), respectively. The diameter of all three oil tanks was 1.5748~cm (4~in). These values were chosen based on the capabilities of the in-house wave tank for experimental testing. Figure~\ref{fig:WaveTank} illustrates a schematic diagram of the wave tank and the prototype inside it. Table~\ref{table:table2} lists the variables for all experiments.

\begin{table*}
	\tiny
	\centering
	\begin{tabular}{cccc|cc}
		\hline
		\multicolumn{4}{c}{Wave Properties} &
		\multicolumn{2}{c}{Oil Pressure Values (kPa)} \\
		\hline
		Depth (cm)& Height (cm)& Periods (s) & \# of cycles &forward motion & backward motion\\
		91.44 & 15.24 & 1, 1.5, 2, 2.5, 3 & 25 & 68.948 - 413.685 & 137.895 - 413.685 \\
		\hline
	\end{tabular}
	\caption{Parameters for the experiments} \label{table:table2}
\end{table*} 

\begin{figure}[h!]
	\centering
	\includegraphics[scale=0.3]{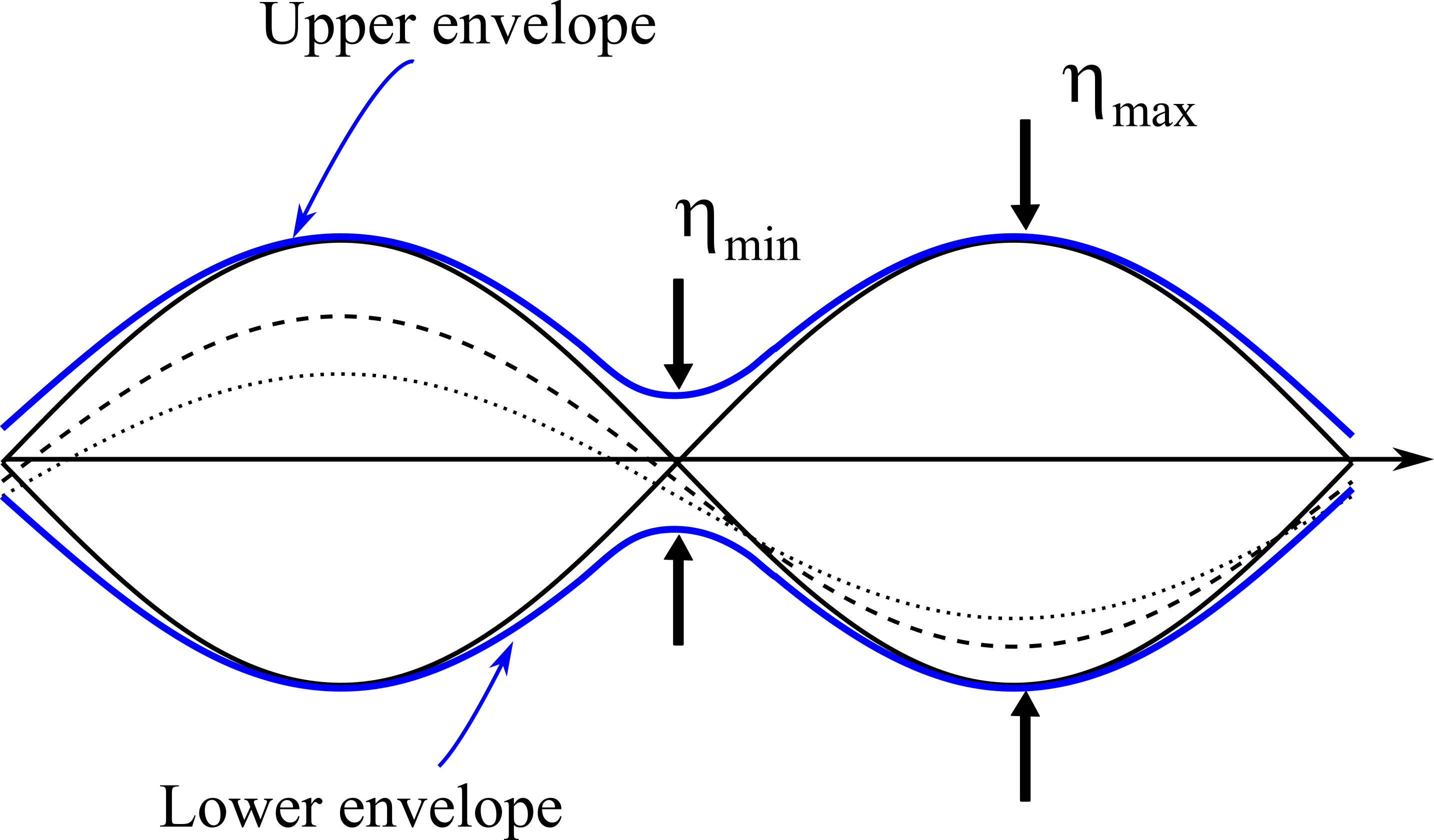}
	\caption{Calculation of reflection and incident waves for the experimental tests.}
	\label{fig:Reflect}
\end{figure}

The incident and reflected wave heights were calculated by measuring the maximum and minimum of the envelope at anti-nodes and nodes~\cite{dean1991water}, as shown in figure~\ref{fig:Reflect}. From these measurements, the height was found using $H_i=\eta_{max}+\eta_{min}$ for the incident waves and using $H_r=\eta_{max}-\eta_{min}$ for the reflected waves.

\subsection{Results \& Discussions}
The displacements of fluid in tanks B and C were measured for 25 cycles of waves. Then, the wave-to-fluid power conversion was calculated based on the measured displacement, the total time, the area of the cylinders, and the pressure values using the equation~\ref{eq:WaveToFluid}. Here, displacement is the change of the fluid level within tanks B and C from the beginning of each experiment to that at its end.

\begin{equation} \label{eq:WaveToFluid}
\small
Power = \frac{work}{time} = \frac{force\times displacement}{time} =\frac{\left(pressure\times area\right)\times displacement}{time}
\end{equation}
\\

\begin{figure}[h!]
	\centering
	\includegraphics[scale=0.55]{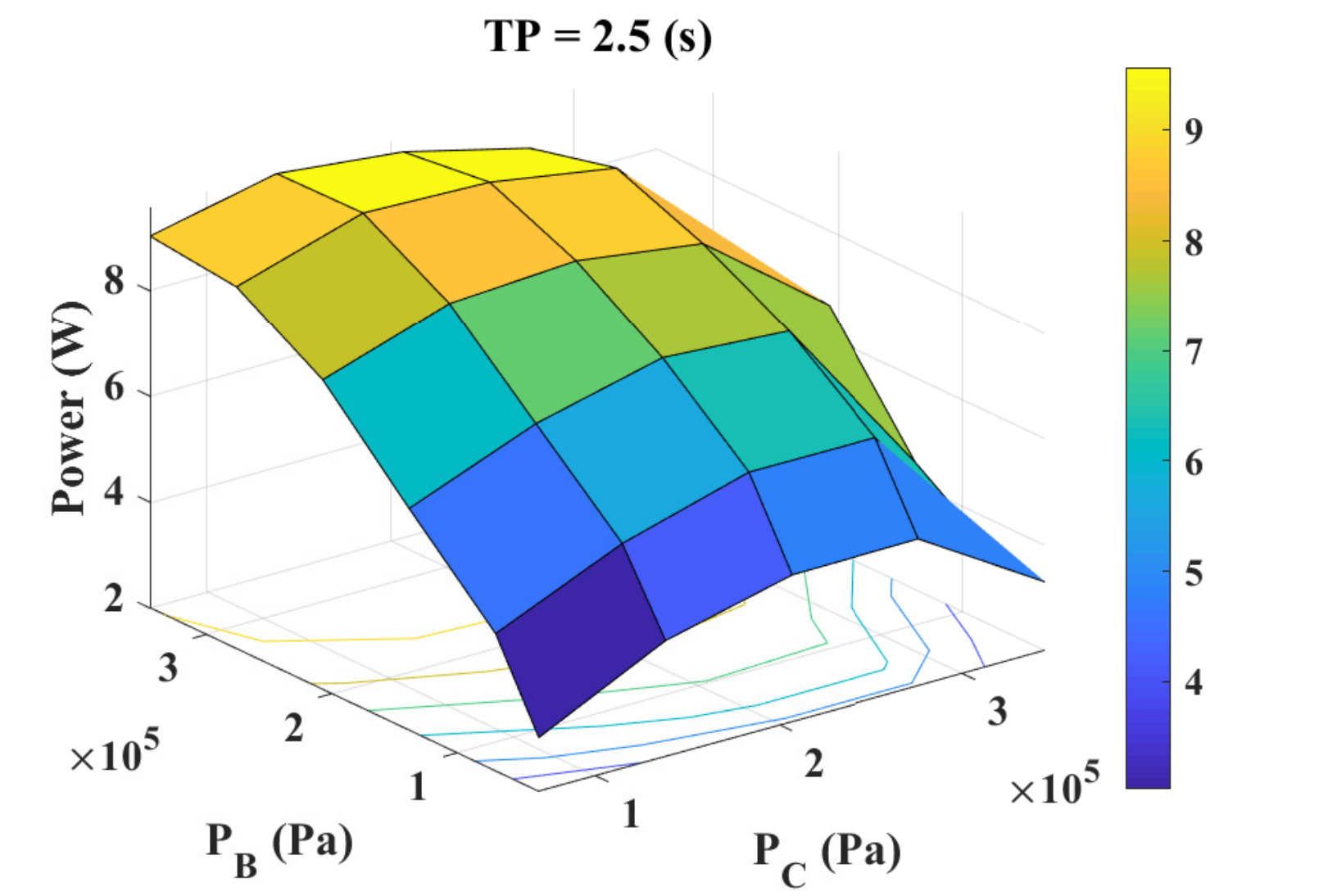}
	\caption{Power extracted after 25 cycles of waves with the 1.5~s time period (TP) and the 15.24~cm height.}
	\label{fig:PowerOutput1.5s}
\end{figure}

Figure~\ref{fig:PowerOutput1.5s} shows the total power for different pressure values. The waves had a time period of 1.5~s and a height of 15.24~cm (6~in).

Based on the results provided in figure~\ref{fig:PowerOutput1.5s}, which are for the waves with a 1.5~s time period, the maximum power of 9.57~W was extracted when $P_C$ = 137.89~kPa and $P_B$ = 275.79~kPa. For low pressure values in tanks B and C, the counter-torque from the PTO system was low, and as a result, the output power was at its minimum. For this condition, $L_A$, which is associated with forward motion, was larger than $L_B$ for the first oscillation. When the mass moment of inertia was not large, there was not enough force to bring back the flap to its initial position for the second oscillation, and therefore, the flap stayed tilted for the rest of its motion. This created a small oscillation around a non-vertical position, causing low stroke lengths, and consequently, a small angular velocity, i.e. $\dot{\theta}(t)$. For higher values of $P_C$, the power output still remained low due to the high damping effect for the backward motion during the oscillation of the flap. However, a small increase was observed compared to low $P_C$ values. This was caused by the fact that the damping effect slightly shifted the natural frequency of the harvester and resulted in a parabolic shape line for each $P_B$ value. Further, the backward motion of the flap was very limited for high values of $P_C$ due to the high pressure in the hydraulic line controlling that motion. Increasing $P_B$ led to much higher power output, which was caused by high counter-torque values, i.e. PTO torque.

\begin{figure}
	\centering
	\includegraphics[scale=.4]{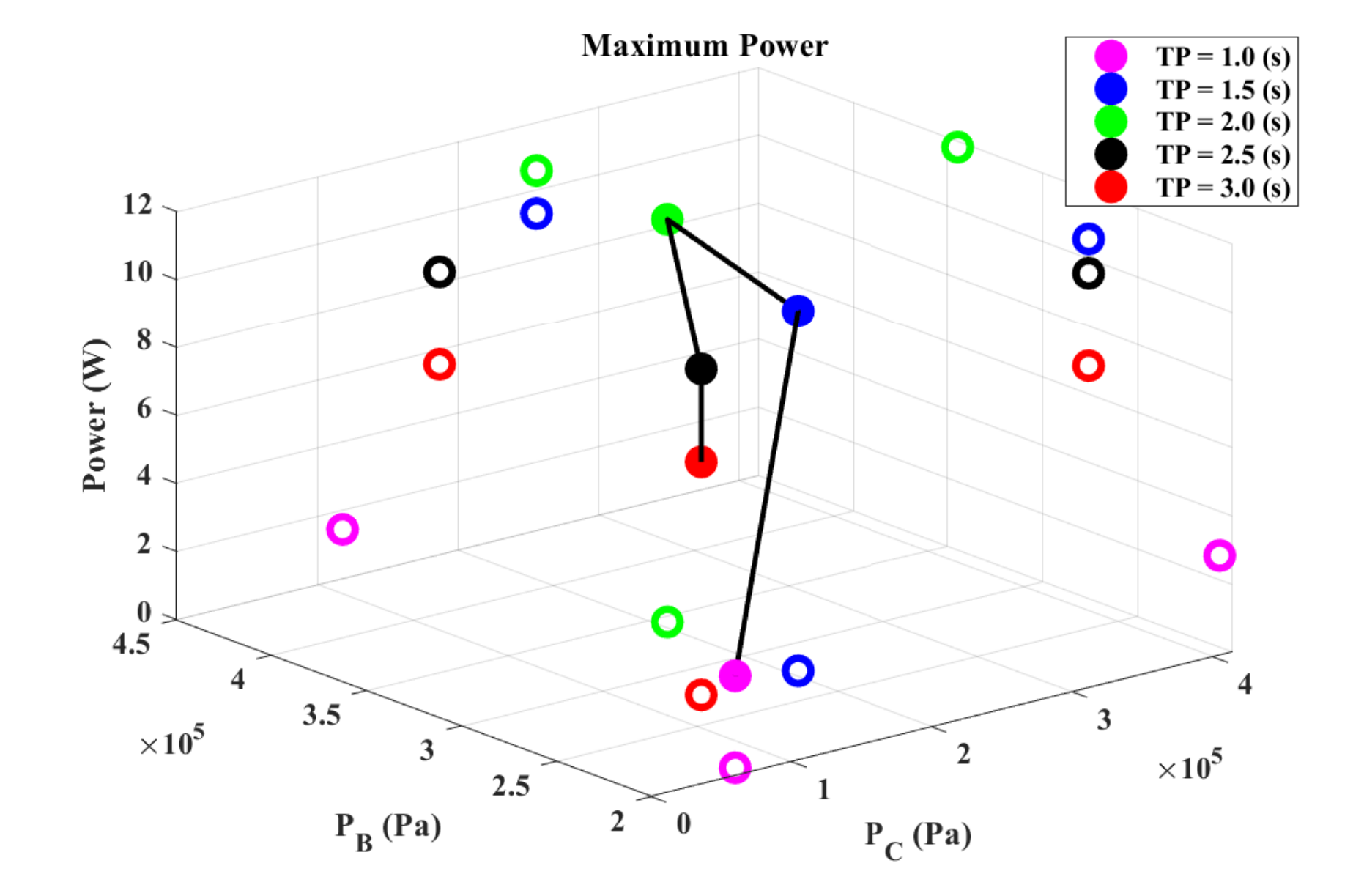}
	\caption{Maximum power extracted after 25 cycles of the waves with 15.24~cm height.}
	\label{fig:MaxPowerOut}
\end{figure}

Similar procedures were followed for the waves with time periods listed in Table~\ref{table:table2} to find the maximum power outputs and their corresponding pressure values in tanks B and C. Results are shown in Fig.~\ref{fig:MaxPowerOut}. In this figure, hollow circles represent the projections of data points on different planes to help identify the location of each point in the three dimensional space. Maximum power output was found at $P_C$ = 206.84~kPa and $P_B$ = 344.74~kPa with the value of 11.84~W.

Based on results provided in figure~\ref{fig:MaxPowerOut}, the maximum power output first increased, and then, decreased by increasing the time period of incident waves. For this prototype, the highest power output is observed for waves with a 2~s time period. Maximum outputs occur at different pressure values for tanks B and C. The highest change in maximum power occurred when the time period was changed from 1~s to 1.5~s. Based on the pattern in Figure~\ref{fig:MaxPowerOut}, it can be seen that the prototype has high power outputs for the waves with time periods 1.5~s to 2.5~s, i.e. frequencies between $\frac{1}{2.5}$~Hz and $\frac{1}{1.5}$~Hz. Once these values were scaled for this 1/20$^{th}$ scale device using Froude’s law (since the inertia forces are predominant), the final flap had the maximum power output in the frequency range between 0.09~Hz and 0.15~Hz. This frequency range for high power output is expected, as these waves have high power density, particularly close to 0.09~Hz. (see Figure~\ref{fig:PowerOutputTankBTP15} for waves' spectral density distribution).

\begin{figure}
	\centering
	\includegraphics[scale=0.25]{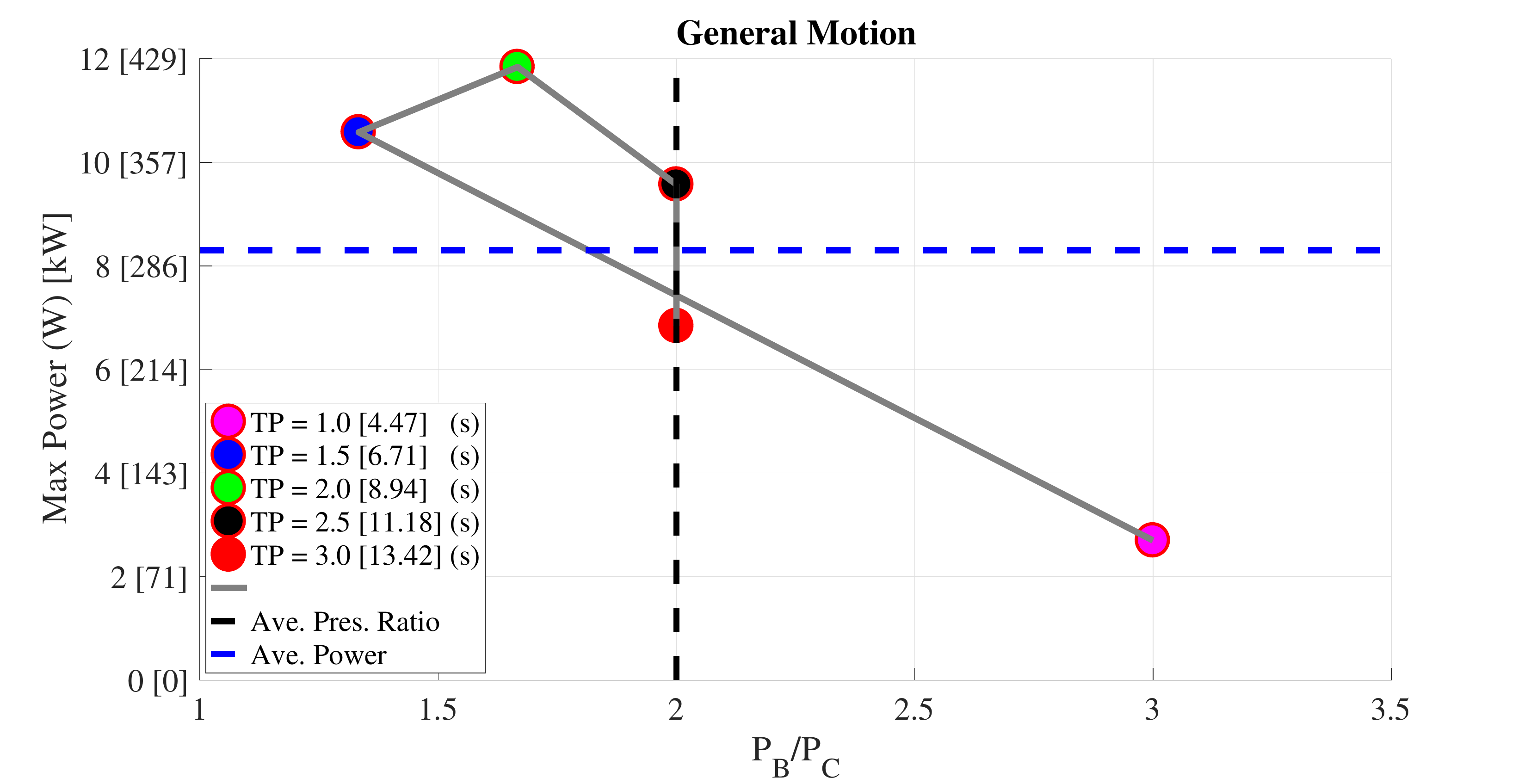}
	\caption{Maximum power extracted after 25 cycles of waves with various time periods at different pressure ratios.}
	\label{fig:MaxPowerOutGeneralMotionPeriods}
\end{figure}

Since the device is designed to have a hydraulic system with only one oil pressure value for the whole circuit, the ratio between forward motion and backward motion pressure values can be found to help with the design process. Once the best pressure ratio for the PTO mechanism is found, the area of both sides of the pistons can be designed to provide desirable forces and torques. Figure~\ref{fig:MaxPowerOutGeneralMotionPeriods} shows the maximum power output values for different wave periods. However, it is represented based on the ratios between pressures in tanks B and C. In this figure, numbers in brackets are the scaled values for the final device with the scaling factor of 20. For this prototype, the average power output was more than 8~W in the face of waves with periods from 1~s to 3~s (blue dashed line in figure~\ref{fig:MaxPowerOutGeneralMotionPeriods}). The average pressure value with the maximum power output for waves in this period range is $\frac{P_B}{P_C}$ = 2. Therefore, a hydraulic circuit with a pressure ratio of 2 between two sides of the pistons can produce an average power of approximately 290~kW for the scaled machine with one flap. For this ratio, the maximum power output corresponds to the waves with a dominant wave period of 11.18~s (0.09~Hz frequency).

\begin{figure}
	\centering
	\includegraphics[scale=0.55]{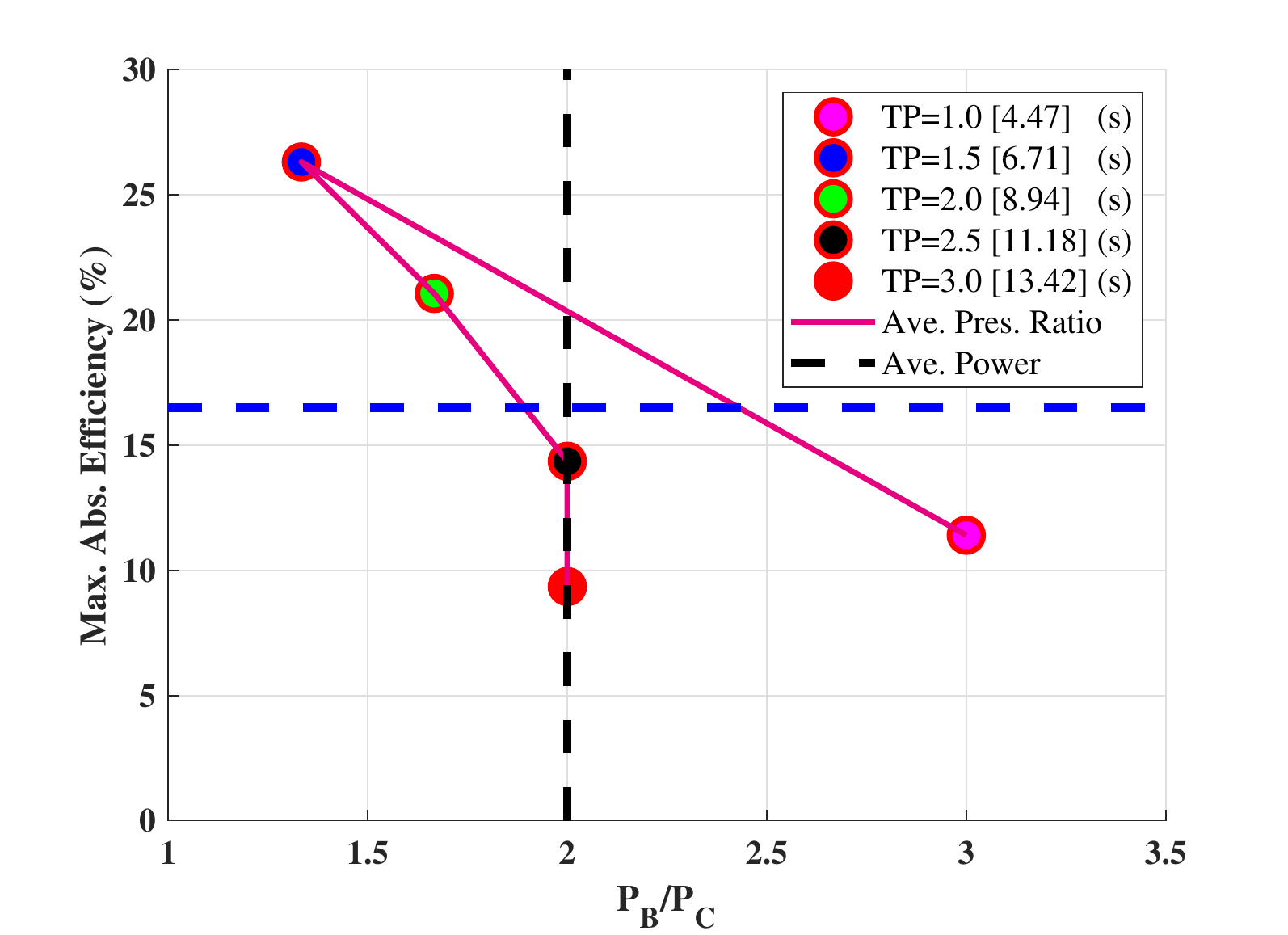}
	\caption{Wave-to-fluid efficiency for different time periods at pressure ratios.}
	\label{fig:efficiency}
\end{figure}

Finally, the efficiency of the harvester can be calculated based on the energy flux of the incident waves and the output power. The energy per unit time and unit width of the wave frontage (or energy flux) can be found using~\cite{goda2010random}

\begin{equation} \label{Eqn:Effi}
J=\frac{\rho g^2D(kh)}{32\pi}TH^2
\end{equation}
where $k$ is the wave number, $T$ is the time period of the wave, and $H$ is the wave height. In equation~\ref{Eqn:Effi}, $D(kh)$ is the depth function defined as: 

\begin{equation} \label{Eqn:DepthFunc}
D(kh)=\left[1-\frac{\omega^2}{gk}\right]kh+\frac{\omega^2}{gk}
\end{equation}
\\
Therefore, the wave-to-fluid efficiency of the device can be calculated. The results of these calculations are shown in Figure~\ref{fig:efficiency}. Although the maximum power output was found for waves with 2~s time period (8.94~s for the scaled machine), the maximum efficiency of the device occurred for waves with a 1.5~s time period (6.71~s for the scaled machine). On average and for this range of wave periods, the capturing efficiency was about 16\%. It is worth mentioning that this efficiency is for a prototype, which is not optimized in terms of its natural frequency. Finally, this device can also be designed for the maximum generated power by changing the design pressure ratio to 1.3 in order to reach a 26\% efficiency. However, it should be installed in the areas with dominant wave periods of 6.71~s.

\section{Remarks \& Conclusions}
In this paper, analytical investigations were used to provide a general working principle of a flap-type seabed-hinged wave energy harvester with a hydraulic PTO. The natural frequency of the device can be controlled using fillable compartments as part of the design. These compartments can also be used for storm protection to sink the flap during stormy weather conditions as a survival mechanism. A series of experimental tests were designed and conducted on a 1/20$^{th}$ scale device to study the effects of damping and PTO torque caused by oil pressure in the hydraulic circuit. Optimal power values for the converter were recorded for waves with time periods varying from 1~s (4.47~s for the scaled device) to 3~s (13.42~s for the scaled device). The maximum power output of 11.84~W occurred at a 2~s time period. To get the maximum power over a range of time periods between 1~s and 3~s, the average ratio between pressures in forward and backward motions should be set to 2. Since the pressure and force are correlated, this value can be used to design the cylinder-piston gadgets. To design the device for a specific site or environment, pressure ratios in the hydraulic circuit and the natural frequency of the device can be optimized. For waves with a height of about 3.048~m (10~ft) in a water depth of 18.288~m (60~ft), this device can reach up to a 26\% water-to-fluid efficiency, for wave periods between 4.47~s-13.42~s with a non-optimized natural frequency.

\section*{Acknowledgments}
The author would like to thank the New York State Research and Development Authority (NYSERDA) and Clean Energy Business Incubator Program (CEBIP) at Stony Brook University for their support.

%\nocite{*}% Show all bib entries - both cited and uncited; comment this line to view only cited bib entries;
%\bibliography{wileyNJD-AMA}%
\bibliographystyle{ieeetr}
\bibliography{References}

\end{document}